\documentclass[lettersize,journal]{IEEEtran}
\usepackage{amsmath,amssymb,amsfonts}
\usepackage[linesnumbered,ruled,vlined]{algorithm2e}
\usepackage[caption=false,font=normalsize,labelfont=sf,textfont=sf]{subfig}
\usepackage{textcomp}
\usepackage{stfloats}
\usepackage{url}
\usepackage{verbatim}
\usepackage{graphicx}
\usepackage{hyperref}
\usepackage{cite}
\usepackage{fancyhdr}
\usepackage{tabularray}
\usepackage[inline]{enumitem}
\usepackage{xcolor}
\usepackage{comment}
\usepackage{diagbox}
\usepackage{tikz}
\usepackage[switch]{lineno}

\begin{document}

\title{Compass: Co-Exploration of Mapping and Hardware for Heterogeneous Multi-Chiplet Accelerators Targeting LLM Inference Service Workloads}

\author{Boyu~Li,
Zongwei~Zhu\textsuperscript{$\ast$},
    Qianyue~Cao,
    Xi~Li, and
 Xuehai~Zhou,~\IEEEmembership{Member,~IEEE, }
 
\thanks{This work was supported by the National Natural Science Foundation of China (62572455). \emph{(Corresponding author: Zongwei Zhu.)}}

\thanks{Boyu Li, Qianyue Cao, and Xuehai Zhou are with the School of Computer Science and Technology, University of Science and Technology of China, Hefei 230026, China and Suzhou Institute for Advanced Research, University of Science and Technology of China, Suzhou 215123, China  (e-mail: llbbyy@mail.ustc.edu.cn; cqy\_1999@mail.ustc.edu.cn; xhzhou@ustc.edu.cn).}

\thanks{Zongwei Zhu and Xi Li are with the School of Software Engineering, University of Science and Technology of China, Hefei 230026, China and Suzhou Institute for Advanced Research, University of Science and Technology of China, Suzhou 215123, China (e-mail: zzw1988@ustc.edu.cn; llxx@ustc.edu.cn
).}

}

\maketitle

\begin{abstract}

Large language models (LLMs) bring huge computational demands, which makes multi-chiplet accelerators that can integrate large-scale computing resources a powerful solution. However, existing design space exploration (DSE) efforts for such accelerators primarily focus on traditional CNN/Transformer workloads and fall short in supporting the highly dynamic behavior of real-world LLM inference services. This dynamic nature manifests in two key aspects: 1) Mixed request types: the prefill and decode phases exhibit significantly different computational patterns and are frequently interleaved by modern system-level service schedulers; 2) Variable sequence lengths: the sequence length differences across requests can span several orders of magnitude, rendering padding-based assumptions inefficient. Moreover, many prior works assume homogeneous chiplets and overlook the potential beneficial interaction between LLM dynamics and heterogeneous chiplet architectures. To bridge this gap, we introduce Compass, a co-exploration framework designed to optimize mapping strategies and hardware design for multi-chiplet accelerators, specifically tailored for dynamic LLM workloads. First, we propose a computation execution graph-based mapping encoding scheme that decouples micro-batch and layer dimensions, enabling fine-grained execution control on heterogeneous chiplets and flexibly representing various parallelism strategies. Second, based on this scheme, we develop the Compass framework itself, which integrates an evaluation engine, a mapping generation engine based on genetic algorithm, and a hardware sampling engine based on Bayesian optimization, enabling fast and flexible cross-level co-design. Compared with the SOTA DSE works Gemini and MOHaM, Compass reduces latency by 63.92\% and energy by 40.32\% on average in various scenarios, with only a 3.11\% increase in monetary cost. Furthermore, we combine Compass with cutting-edge inference service scheduling strategies such as Chunked Prefill, demonstrating the mutual influence between scheduling policies and multi-chiplet accelerators.

\end{abstract}


\section{Introduction}\label{sec:intro}

In recent years, large language models (LLMs) have emerged as a central focus in the field of deep learning, demonstrating remarkable capabilities across a wide range of natural language processing tasks. Corresponding to their outstanding performance is a tremendous demand for computation and memory, making deep learning accelerators a hot research topic. However, due to process limitations and chip yield constraints, the performance of single-chip accelerators has gradually reached its limit\cite{single-chip-limit}. Multi-chiplet accelerators have been proposed as a potential solution to this problem\cite{many_chip_gpu,big-little-chiplet}. These accelerators leverage advanced packaging technologies to integrate multiple chiplets, constructing large-scale acceleration systems.

\begin{figure}[ht]
\centering
\includegraphics[width=0.48\textwidth]{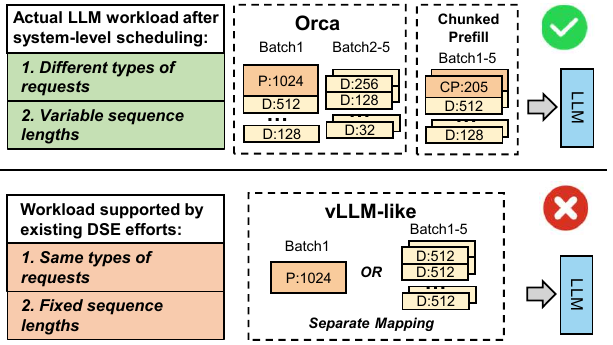}
\caption{The differences between the workload characteristics in LLM inference service scenarios and those supported by existing DSE frameworks are mainly reflected in two aspects: mixed request types and variable sequence lengths.}
\label{fig:intro}
\end{figure}

Compared to monolithic systems, multi-chiplet accelerators introduce novel challenges, including Network-on-Package (NoP) communication bottlenecks, chiplet cost and reusability considerations, and layer-pipeline modeling. Consequently, this has catalyzed a proliferation of works on mapping and hardware design space exploration (DSE) \cite{gemini,moham,scar}. However, most existing efforts are confined to traditional deep learning models (such as CNNs and standard Transformers). These works treat LLMs as a special case of conventional DNNs and consistently explore design schemes targeting individual request types  (either prefill or decode) with specified sequence lengths. In reality, LLM workloads in inference serving scenarios are highly variable. Coupled with the emergence of system-level scheduling optimizations like Orca, the computational patterns faced at the hardware level during inference serving differ significantly from those in traditional deep learning, exhibiting profound dynamicity. This dynamicity is primarily manifested in two aspects: \textbf{mixed request types} and \textbf{variable sequence lengths}.

\textbf{Mixed Request Types}: Fig. \ref{fig:intro} illustrates the workload orchestration in two state-of-the-art inference serving scenarios: Orca \cite{orca} and Chunked Prefill \cite{chunked_prefill_osdi}. Orca co-executes prefill requests together with decode requests upon their arrival, while Chunked Prefill partitions prefill requests into multiple chunks and executes them jointly with decode requests. However, existing works cannot support such scenarios where requests with different computational behaviors are executed together within the same batch.

\textbf{Variable Sequence Lengths}: LLM serving faces extreme sequence length variability (e.g., 1-161,281 in ShareGPT \cite{sharegpt}). This not only causes different computational behaviors among instances within a batch, but also leads to significant workload variations across batches. Therefore, it is necessary to identify optimal mapping schemes based on the sequence length distribution of the serving scenario. However, existing works typically assume one specific sequence length or a fixed set of sequence lengths.

Existing multi-chiplet accelerator DSE works can be categorized into two classes based on their target workloads: targeting a single model \cite{gemini,stream} and targeting multiple models simultaneously \cite{moham,scar}. However, both classes face challenges in handling LLM inference.

Single-model DSE works assume identical computation and memory access patterns across different instances within a batch, and extrapolate to the entire inference cycle by computing latency and energy for a steady state. For instance, SET\cite{set} and Gemini \cite{gemini} treat a single micro-batch as the steady state, while Stream \cite{stream} captures the steady state during runtime. The theoretical foundations of these approaches can only model LLMs with uniform types and sequence lengths, making it difficult to support more dynamic scenarios through minor framework adjustments.

Multi-model DSE works primarily face two main issues. First, they are not specifically designed for LLMs. If forced to adapt, they can only treat different instances within a batch as independent models for mapping. However, to increase computational intensity, actual LLM inference merges all requests during the QKV generation and FFN phases, splitting them by request only during the attention computation phase. The independent mapping assumption of these frameworks is fundamentally incompatible with the dynamic execution patterns of LLMs, which severely restricts their explorable mapping space. Second, existing multi-model DSE works typically rely on coarse-grained performance modeling, lacking fine-grained tiling analysis and support for layer-pipelining. This is also the reason why previous multi-model mapping works cannot replace single-model DSE works. Therefore, there is a critical need for an LLM-specific DSE framework that provides fine-grained modeling comparable to single-model DSE works.

\definecolor{cellgreen}{HTML}{E5FFE5}
\definecolor{cellred}{HTML}{FFE5E5}
\begin{table}
\centering
\caption{EDP ratio (OS / WS) across computation phases and sequence lengths on GPT3-7B. (\colorbox[HTML]{E5FFE5}{\phantom{X}}: WS superior, \colorbox[HTML]{FFE5E5}{\phantom{X}}: OS superior)}
\label{tab:edp compare for diff chip}
\scalebox{0.9}{
\begin{tblr}{
  width = \linewidth,
  colspec = {Q[173]Q[240]Q[162]Q[156]Q[156]},
  cells = {c},
  hlines,
  vlines,
}
  \diagbox[width=5.5em, height=2.5em]{Lens}{Phase} & QKV Gen & QK$^{T}$ & FFN1 & FFN2 \\
128   & \SetCell{bg=cellgreen}3.35x & \SetCell{bg=cellgreen}1.32x & \SetCell{bg=cellgreen}2.43x & \SetCell{bg=cellgreen}3.38x \\
1024  & \SetCell{bg=cellgreen}2.43x & \SetCell{bg=cellred}0.88x   & \SetCell{bg=cellgreen}2.46x & \SetCell{bg=cellgreen}2.45x \\
5120  & \SetCell{bg=cellred}0.96x   & \SetCell{bg=cellred}0.33x   & \SetCell{bg=cellred}0.85x   & \SetCell{bg=cellgreen}1.79x \\
10240 & \SetCell{bg=cellred}0.84x   & \SetCell{bg=cellred}0.31x   & \SetCell{bg=cellred}0.85x   & \SetCell{bg=cellred}0.85x  \\
\end{tblr}
}
\end{table}

Beyond the inherent characteristics of LLMs, \textbf{the interplay between LLM dynamism and chiplet heterogeneity also presents an interesting topic}. Table \ref{tab:edp compare for diff chip} presents the energy-delay product (EDP) ratios of general matrix multiplication (GEMM) across different phases of GPT3-7B under varying input sequence lengths, comparing output-stationary (OS) chiplets to weight-stationary (WS) chiplets. It can be observed that changes in sequence length and phases of LLM inference produce different dataflow type preferences. This inspires us to use a heterogeneous chiplet architecture, enabling us to allocate GEMM workloads with different preferences to the corresponding chiplets for execution. Existing works such as Gemini focus on DSE for homogeneous multi-chiplet accelerators and do not support heterogeneous chiplets.

In summary, LLM workloads differ significantly from traditional deep learning workloads in both intrinsic behavior and their interaction with chiplet architectures. Yet, prior DSE efforts\cite{gemini,moham,scar} for multi-chiplet accelerators neither support such LLM execution patterns nor thoroughly investigate mapping strategies and corresponding chiplet design strategies tailored for LLMs. To address these challenges, we make the following contributions:

\begin{enumerate*}[itemjoin=\\\hspace*{\parindent}]
    \item We propose a computation execution graph-based mapping encoding scheme. This scheme supports fine-grained scheduling of the minimal dynamic units in LLMs by decoupling micro-batches and layers. It can flexibly represent various mapping strategies (such as model, data, and pipeline parallelism on chiplets), providing a foundational representation scheme for subsequent mapping and hardware exploration. To our knowledge, this is the first work to systematically define the mapping space for dynamic LLM inference workloads.

    \item Building upon the encoding scheme, we develop Compass, an accelerator DSE framework supporting heterogeneous chiplet designs and dynamic LLM workloads. The evaluation engine is the core of Compass. Based on a fine-grained data access analysis algorithm, it can evaluate the latency, energy consumption, and monetary cost under a given hardware and sequence length distribution. Furthermore, Compass integrates a genetic algorithm-based mapping search engine and a Bayesian optimization-based hardware sampling engine to enable efficient design space exploration. \textbf{Compass is open-sourced at: \url{https://anonymous.4open.science/r/Compass-CB4A}}.
    
    \item Compared with the state-of-the-art (SOTA) DSE works Gemini and MOHaM, the solutions provided by Compass reduce latency by 63.92\% and energy by 40.32\% on average in various scenarios, with only a 3.11\% increase in monetary cost. It demonstrates the significant benefits of considering LLM dynamism during the DSE process. In addition, we also analyze its interaction with cutting-edge inference service scheduling strategies such as Chunked Prefill, showcasing the mutual impacts between these strategies and multi-chiplet accelerators. We hope these case studies can inspire researchers and promote the co-optimization of multi-chiplet accelerators and dynamic LLM workloads.
\end{enumerate*}

\section{Related Work}

\textbf{LLM Inference Service Optimization.} Numerous system-level efforts have emerged to accelerate LLM inference services since the advent of LLMs. Among them, Orca \cite{orca} is the first to propose iteration-level scheduling during batching: batching is applied during non-multi-head attention (MHA) stages, while requests are handled individually during the MHA stage. This eliminated the need for static batching \cite{fastertransformer}, enabling mixed requests with different types and sequence lengths in a single batch. vLLM \cite{vllm} also uses iteration-level scheduling but pauses ongoing decode processing upon the arrival of a prefill request, prioritizing prefill execution. From a chip-level perspective, this results in a type-separated workload pattern. Chunked Prefill \cite{chunked_prefill_osdi} introduces a strategy that splits prefill requests into multiple chunks and interleaves them with decode requests. These works have dramatically changed the nature of workloads faced by accelerators compared to traditional neural networks, leading to new challenges around mixed request types and variable sequence lengths.

\newcommand{\checkedwithslash}{%
  \begin{tikzpicture}[baseline=-0.6ex]
    \node at (0,0) {$\checkmark$};
    \draw[thick] (-0.3em, 0.3em) -- (0.4em, -0.3em);
  \end{tikzpicture}%
}
\begin{table}
\centering
\caption{Comparison with Related DSE Works}
\label{tab:related work}
\scalebox{0.9}{
\begin{tblr}{
  width = \linewidth,
  colspec = {Q[240]Q[167]Q[170]Q[175]Q[200]},
  cells = {c},
  columns = {m},
  vline{2} = {-}{},
  hline{1,10} = {-}{0.08em},
  hline{2} = {-}{0.05em},
  hline{6,9} = {-}{},
}
\textbf{Work}   & \textbf{LLM Dynamism}~~    & \textbf{Hardware DSE}~~ & \textbf{Pipelining Modeling}~~ & \textbf{Evaluation} \\
Simba\cite{simba}  &                   &                & $\checkmark$                        & L\&E         \\
SET\cite{set}    &                   &                & $\checkmark$                        & L\&E         \\
Gemini\cite{gemini} &                   & Homo           & $\checkmark$                        & L\&E\&MC       \\
Stream\cite{stream} &                   & Hetero         & $\checkmark$                        & L\&E         \\
MAGMA\cite{magma}  &  &                &                                   & L\&E         \\
MOHaM\cite{moham}  &  & Hetero         &                                   & L\&E\&Area        \\
SCAR\cite{scar}   &  &                & $\checkmark$                        & L\&E         \\
\textbf{Ours}   & $\checkmark$           & Hetero         & $\checkmark$                        & L\&E\&MC       
\end{tblr}
}
\end{table}


\textbf{DSE for Chiplet Accelerators.}  Table \ref{tab:related work} presents a comparison between our work and prior DSE works. As discussed in Section \ref{sec:intro}, related works can be divided into two groups. Works targeting single-model typically assume identical computation patterns across instances, and thus cannot support LLM dynamism. Works targeting multiple models also fall short, as their independent execution assumption conflicts with the merge-split execution pattern in LLM inference, and many lack fine-grained performance modeling for layer-pipelining. In summary, our work is specifically designed for dynamic LLM workloads, incorporates heterogeneous chiplet DSE, and provides multi-metric evaluation covering latency (L), energy (E), and hardware monetary cost (MC), grounded in fine-grained data access analysis of layer-pipelining.

\textbf{Wafer-Scale Systems.} Recently, works such as WSC-LLM have explored the co-design of LLM serving and architecture for wafer-scale systems \cite{wsc-tcad,wsc-llm}. However, wafer-scale systems are essentially cluster-scale computing architectures. Each compute node within them is typically a complete chip equipped with independent DRAM; therefore, such works primarily focus on deploying different service instances of models and executing distributed parallelism. In contrast, this paper and previous multi-chiplet accelerator research target single-card-level accelerator designs. Under this architecture, each chiplet is typically a compute die relying solely on limited on-chip SRAM. Consequently, our work focuses on the fine-grained partitioning of layers within a single model and the orchestration of efficient layer-pipelining across chiplets. The model slices allocated to compute nodes in wafer-scale research can still leverage our work for further mapping optimization. In summary, this work and wafer-scale-related studies differ in system hierarchy and scheduling granularity, belonging to orthogonal research spaces.

\section{Background}

\subsection{LLM Inference Process}\label{sec:llm process}

\begin{figure}[ht]
\centering
\includegraphics[width=0.49\textwidth]{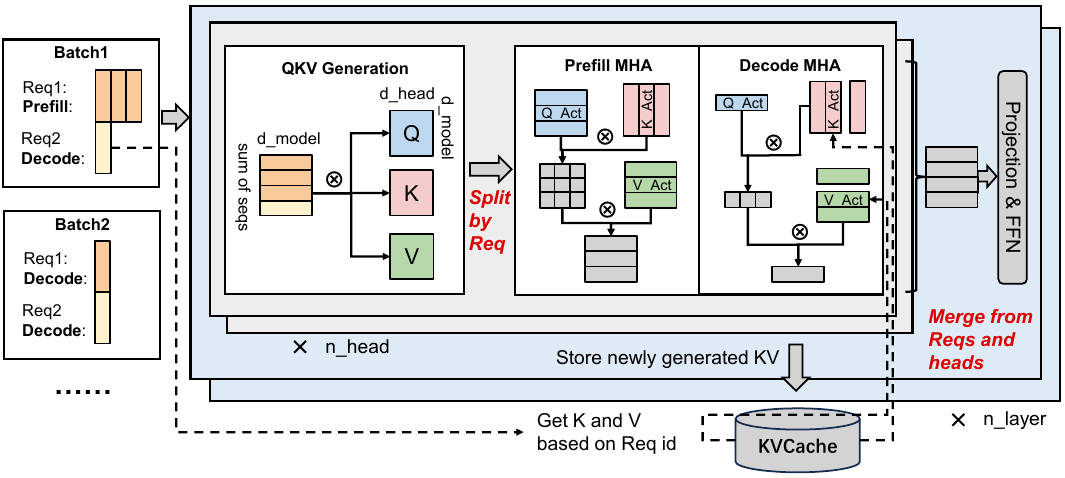}
\caption{The LLM inference process in inference service scenarios. Compared with a traditional transformer, the most distinctive feature lies in the \textbf{splitting and merging operations} caused by variable sequence lengths and mixed request types.}
\label{fig:background_llm}
\end{figure}

Fig. \ref{fig:background_llm} illustrates the LLM workload execution process in inference serving when combined with dynamic batching paradigms. First, the workload in a batch is composed of multiple requests, each potentially differing in type and sequence length. Workloads across different batches may also vary. During inference, the workload of a batch is first \textbf{merged} into a large matrix containing all sequence lengths within the batch for QKV generation, in order to increase computational intensity. Subsequently, due to differences in request types and sequence lengths, these activations are \textbf{split} according to the input requests and processed individually. After multi-head attention (MHA) computation completes, the computational results from different requests are \textbf{merged} into a single matrix for subsequent FFN layer computation. This process of merging, splitting, and re-merging data within a batch distinguishes LLM workload execution from traditional neural networks.

\subsection{Chiplet Hardware Template}

\begin{figure}[ht]
\centering
\includegraphics[width=0.48\textwidth]{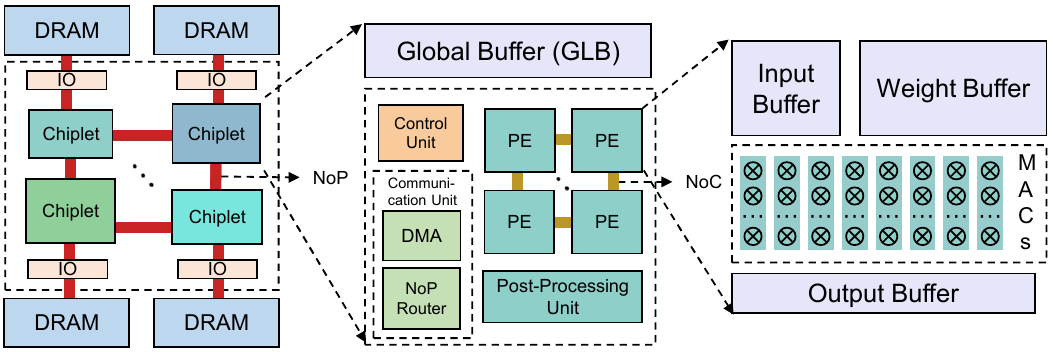}
\caption{Multi-chiplet accelerator hardware template.}
\label{fig:background_hardware}
\end{figure}

Fig. \ref{fig:background_hardware} introduces the hardware template used in this study, derived by abstracting common features from existing chiplet accelerators\cite{magnifier,monad,m2m}. Multiple compute chiplets are interconnected via a Network-on-Package (NoP) to enable inter-chiplet communication. Edge chiplets are connected to IO dies for off-chip DRAM communication. The compute chiplets can be heterogeneous with different dataflows.

Each chiplet includes a Global Buffer (GLB), post-processing units, control units, communication units, and a PE array. The GLB stores input/output activations and weights to support data reuse after tiling. Post-processing units consist of vector processors responsible for scalar operations and nonlinear functions (e.g., activation and normalization) following GEMM or GEMV computations. The control unit generates control signals for internal chiplet operations. The communication unit, comprising DMA and NoP routers, handles data exchange between chiplets and with DRAM.

The PE array consists of compute cores interconnected by an on-chip network, each with input, weight, and output buffers along with a MAC array. These buffers allow further partitioning of GLB data to enhance energy efficiency through data reuse. The MAC array handles parallel processing of computational workloads.

By tuning various configuration parameters of this template, we can emulate a wide range of existing chiplet accelerators.

\section{Mapping Encoding Scheme}\label{sec:mapping encoding scheme}

In this section, we present the proposed mapping encoding scheme. This scheme describes how LLM workloads are mapped onto chiplets and executed under fine-grained control. At the end of this section, we demonstrate how the proposed encoding scheme can represent data parallelism, model parallelism\cite{hybrid-parall}, and pipeline parallelism\cite{gpipe} of workloads across chiplets to illustrate its flexibility.

As mentioned in Section \ref{sec:llm process}, each instance in an LLM workload batch handles different tasks. Therefore, we treat the LLM workload as a two-dimensional computation execution graph \cite{flexgen} with micro-batch and layer as dimensions, rather than the one-dimensional layer sequence assumed in prior work \cite{scar,epipe,set,tangram}. Consider a workload with batch size $N$ and $M$ layers to be mapped onto an accelerator with $C$ chiplets. The encoding scheme consists of three components: $micro\_batch\_size$, $segmentation$, and $layer\_to\_chip$.

$micro\_batch\_size$ is an integer that describes how the computation graph is divided along the micro-batch dimension. It must satisfy $ micro\_batch\_size \mid N$.

$segmentation$ is a binary vector of length $M-1$ that indicates how the graph is segmented along the layer dimension. Formally, $segmentation[i] = 1$ means a segment boundary is placed after layer $i$; $0$ means no segmentation. Since a natural segmentation occurs after the final layer, the length is $M-1$.

After applying segmentation to both dimensions, the original computation graph is divided into $\frac{N\times M}{micro\_batch\_size}$ subgraphs. $layer\allowbreak\_to\_chip$ is a matrix of size $\frac{N\times M}{micro\_batch\_size}$, where each element is an integer between $0$ and $C-1$, indicating which chiplet a subgraph is assigned to for execution.

\begin{figure}[ht]
\centering
\includegraphics[width=0.48\textwidth]{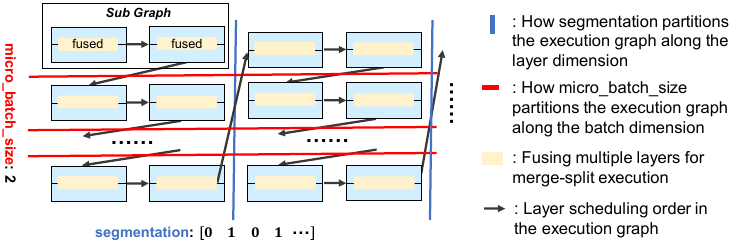}
\caption{State and scheduling order of the computation execution graph after being partitioned into subgraphs.}
\label{fig:encoding_order}
\end{figure}

Fig. \ref{fig:encoding_order} shows the state of the computation execution graph after segmentation and the scheduling order. As described in Section \ref{sec:llm process}, LLM inference typically merges all layers within a batch into a single matrix for processing, except during MHA computation. Therefore, workloads for the same layer within a subgraph are fused. After fusion, tasks within a subgraph are scheduled to their corresponding chiplets in layer order. Then, the subgraphs are scheduled first in micro-batch order, and then in layer order. It is important to note that “scheduling” here refers to the order of assigning workloads to chiplets, not the actual execution order. A workload can be executed only after its predecessor dependencies are resolved and the assigned chiplet becomes available. By leveraging the scheduling order and configuring $segmentation$, the workload can be flexibly scheduled. For example, setting the $segmentation$ to all zeros enables row-wise (layer-first) scheduling; setting it to all ones enables column-wise (micro-batch-first) scheduling; other values of $segmentation$ effectively segment the model itself. This scheduling order allows us to implement various common scheduling strategies by adjusting the values of $micro\_batch\_size$ and $segmentation$.

\begin{figure*}[ht]
\centering
\includegraphics[width=0.98\textwidth]{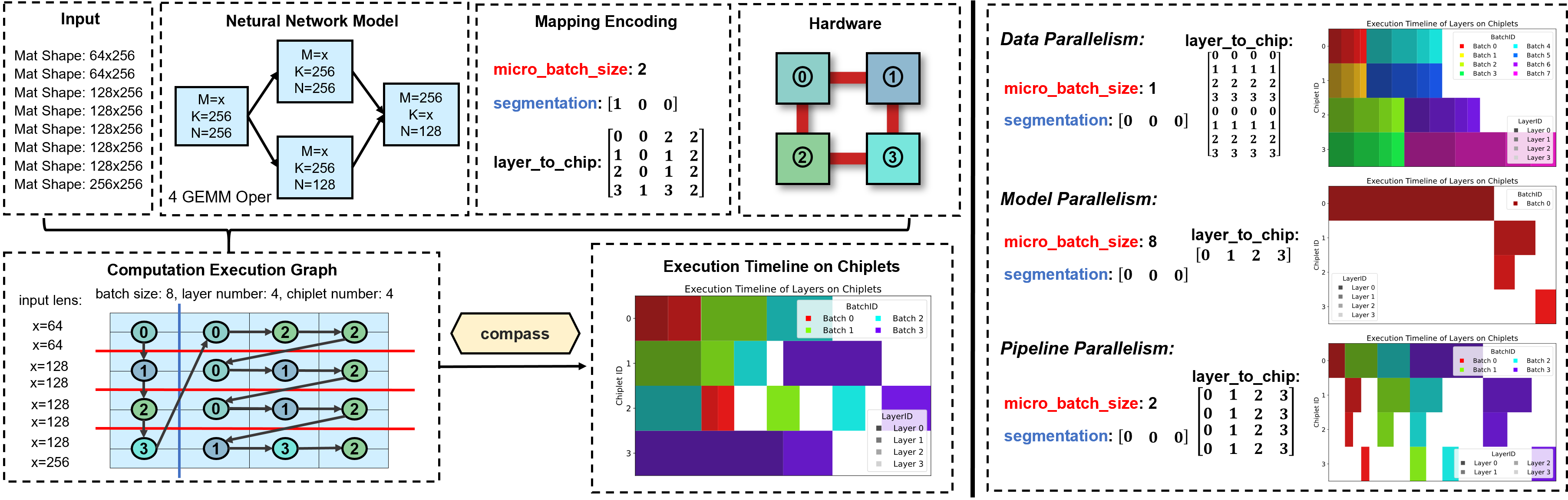}
\caption{An example of a mapping encoding. The workload in this example is a neural network with 4 GEMMs, an input batch size of 8, and variable-length sequence lengths. This workload is to be mapped onto an accelerator with 4 chiplets. The mapping encoding provides a feasible mapping scheme. The lower-left portion shows the computation execution graph corresponding to the mapping encoding and the spatio-temporal diagram of the actual execution process. The right portion presents the mapping encoding representations and spatio-temporal execution diagrams of three common parallelism paradigms.}
\label{fig:encoding_instance}
\end{figure*}

Fig. \ref{fig:encoding_instance} shows a specific example of encoding and scheduling. In this example, the neural network model has four layers, each performing a matrix multiplication of $M\times K$ and $K\times N$. The model processes a batch containing inputs of different matrix sizes. The workload is to be mapped to hardware with four chiplets. The mapping scheme shown in Fig. \ref{fig:encoding_instance} is one found by Compass. The corresponding computation execution graph and the spatio-temporal execution diagram are also illustrated in the figure.

\begin{algorithm}[tb]
\caption{The Mapping Encoding Representation of Three Common Parallelism}\label{alg:parallel inference schemes}
\small
\SetAlgoLined
\DontPrintSemicolon
\KwIn{Total batch size $B$, number of layers $L$, number of chiplets $C$}
\KwOut{$micro\_batch\_size$, $segmentation$, $layer\_to\_chip$}

\textbf{\underline{Data Parallelism:}} \\
$micro\_batch\_size \gets 1$ \\
$segmention \gets$ \texttt{vector}($L-1$), initialized with 0 \\
$layer\_to\_chip \gets$ \texttt{matrix}($B,L$) \\
\For{$i \gets 0$ \KwTo $B-1$}{
    \For{$j \gets 0$ \KwTo $L-1$}{
        $layer\_to\_chip[i,j] \gets i\mod C$\;
    }
}

\textbf{\underline{Model Parallelism:}} \\
$micro\_batch\_size \gets B$ \\
$segmention \gets$ \texttt{vector}($L-1$), initialized with 0 \\
$layer\_to\_chip \gets$ \texttt{matrix}($1,L$) \\
\For{$i \gets 0$ \KwTo $L-1$}{
    $layer\_to\_chip[0,i] \gets i\mod C$\;
}

\textbf{\underline{Pipeline Parallelism:}} \\

$micro\_batch\_size \gets k$ \tcp*{where $1 \leq k \leq B$ and $B \mod k = 0$} 
$segmention \gets$ \texttt{vector}($L-1$), initialized with 0 \\
$layer\_to\_chip \gets$ \texttt{matrix}($B/k,L$) \\
\For{$i \gets 0$ \KwTo $L-2$}{
    \If{$(i+1)\mod C = 0$}{
        $segmention[i] \gets 1$
    }
}
\For{$j \gets 0$ \KwTo $L-1$}{
    \For{$i \gets 0$ \KwTo $B/k-1$}{
        $layer\_to\_chip[i,j] \gets j\mod C$\;
    }
}
\end{algorithm}

This mapping encoding representation can also be used to represent various common parallel inference strategies. Algorithm \ref{alg:parallel inference schemes} describes the mapping encoding configurations for data parallelism, model parallelism, and pipeline parallelism. In data parallelism, the model is split along the batch dimension, with each chiplet independently executing all layers of a batch. There is no inter-chiplet communication, and inter-layer activations can be kept on-chiplet for reuse. In model parallelism, all layers with the same layer ID within a batch are fused together, and the entire model is split by layers and mapped to different chiplets. Inter-layer activations can be transferred via NoP to avoid DRAM accesses. In pipeline parallelism, each chiplet is fixed to process one specific layer and executes batches in sequence like a pipeline, reducing both the DRAM accesses for inter-layer activations and the idle time of chiplets caused by layer dependencies. The right half of Fig. \ref{fig:encoding_instance} shows the specific mapping encoding values and the corresponding spatio-temporal execution diagrams for the three parallel schemes, based on the example in the left half. As shown, these common parallelism strategies are all special cases of the proposed encoding scheme, demonstrating its flexibility. The mapping scheme here is similar to prior works mapping multiple models \cite{scar,magma}, focusing on scenarios where individual layers are mapped to individual chiplets. For tensor parallelism, it can be achieved by configuring layer partitioning in the model architecture, thereby enabling exploration of hybrid parallelism schemes.

\section{Compass Framework}

\begin{figure*}[ht]
\centering
\includegraphics[width=0.98\textwidth]{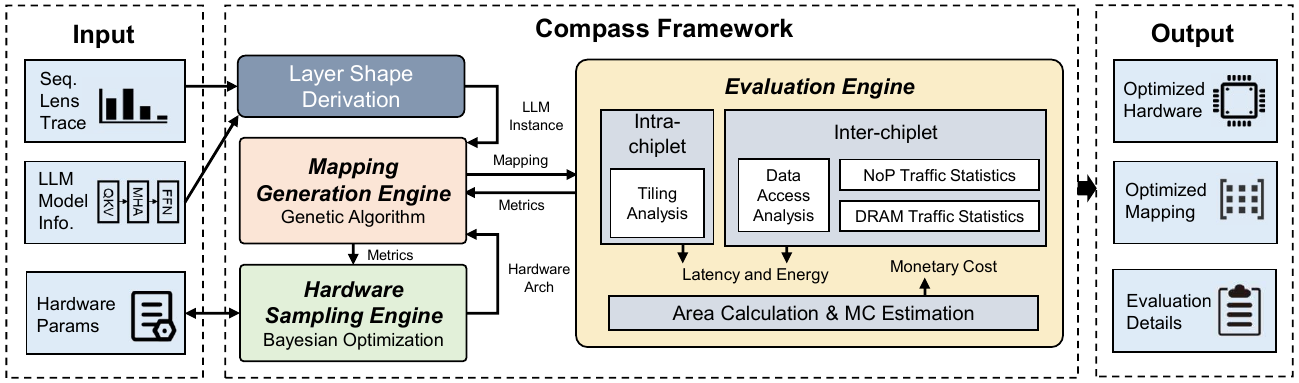}
\caption{Overview of the Compass framework.}
\label{fig:compass}
\end{figure*}

As illustrated in Fig. \ref{fig:compass}, Compass accepts sequence length traces, LLM model information, and hardware candidate parameters as inputs, and produces optimized mapping schemes, hardware architectures, and evaluation details as outputs. The sequence length trace represents a novel input introduced in Compass relative to prior DSE works, enabling mapping generation conditioned on the distribution of sequence lengths. This facilitates workload-specific optimization tailored to distinct application scenarios—for instance, summarization workloads characterized by long inputs and short outputs, or conversational workloads with short inputs and long outputs. Compass is a design-time DSE tool, used to generate sequence distribution-aware hardware designs and mapping schemes during compilation, which is more in line with mainstream NPU graph compilation practices. Although the mapping representation provided by Compass can also support runtime execution adjustments, how to perform efficient dynamic scheduling at runtime is not the problem this paper aims to solve, and is orthogonal to Compass.

Based on this, the problem can be formalized as follows: given a request distribution $\mathcal{D}$, the engine samples a set of requests $\lambda \sim \mathcal{D}$. For the set of candidate hardware architectures $\mathcal{H}$, the set of mapping strategies $\mathcal{M}$, and the accelerator performance-cost function 
$\mathcal{C}$, the goal of Compass is to select the optimal hardware $H^*$ and mapping strategy $M^*$ under the request distribution $\mathcal{D}$ to minimize the expected cost:

\begin{equation}
(H^*, M^*) = \mathop{\arg\min}_{H \in \mathcal{H}, \; M \in \mathcal{M}} \; \mathbb{E}_{\lambda \sim \mathcal{D}} \left[ \mathcal{C}(\lambda, H, M) \right]
\end{equation}

Compass consists of three primary components: a mapping generation engine, a hardware sampling engine, and an evaluation engine. First, Compass determines the layer shapes of each LLM instance within the batch based on the sequence length traces and the LLM model information. The hardware sampling engine then provides the architectural parameters of the currently sampled hardware candidate. The instantiated LLM batch and hardware architecture are jointly fed into the mapping generation engine, which searches for an optimal mapping scheme by iteratively invoking the evaluation engine to obtain latency, energy consumption, and hardware monetary cost estimates for each (LLM, hardware, mapping) triplet. The optimal mapping scheme identified by the mapping generation engine serves as the performance indicator for the corresponding hardware architecture, thereby guiding subsequent hardware sampling iterations until a predefined number of search rounds or a target performance criterion is reached.

Throughout the search process, Compass generates multiple batches from the input traces to capture the average performance across the sequence length distribution. Furthermore, Compass supports additional functionalities including fixed prefill lengths, fixed request-type ratios, and multi-batch generation. These features enable the engine to accommodate scenarios such as Chunked Prefill, facilitating the evaluation of diverse LLM scheduling strategies at the system level.

\subsection{Mapping Generation Engine}

\begin{table}
\centering
\caption{Mutation operators designed for $layer\_to\_chip$. \colorbox[HTML]{E5FFE5}{\phantom{X}}: minimal impact, layer-level; \colorbox[HTML]{FEFDE9}{\phantom{X}}: moderate impact, subgraph-level; \colorbox[HTML]{FFE5E5}{\phantom{X}}: maximal impact, graph-level}
\label{tab:mutation_types}
\begin{tblr}{
  width = \linewidth,
  colspec = {Q[37]Q[906]},
  columns = {m},
  row{1} = {c},
  cell{2}{1} = {c},
  cell{3}{1} = {c},
  cell{4}{1} = {c},
  cell{5}{1} = {c},
  cell{6}{1} = {c},
  cell{7}{1} = {c},
  cell{8}{1} = {c},
    row{7-8} = {bg=red!10},   
  row{5-6} = {bg=yellow!10},
  row{2-4} = {bg=green!10}, 
  hlines,
  vline{2} = {-}{},
  hline{1,9} = {-}{0.08em},
  vline{1,3} = {-}{0.08em},
}
\textbf{ID} & \textbf{Description}                                                                                            \\
1           & Replace one position in $layer\_to\_chip$ with a new random chiplet                                           \\
2           & Swap the chiplet ID of one position in $layer\_to\_chip$ with its adjacent position along the layer dimension \\
3           & Swap the chiplet ID of one position in $layer\_to\_chip$ with its adjacent position along the batch dimension \\
4           & Randomly permute the $layer\_to\_chip$ entries corresponding to a specific subgraph                           \\
5           & Replace every position in the $layer\_to\_chip$ of a specific subgraph with new random chiplets               \\
6           & Swap the $layer\_to\_chip$ entries of one column of subgraphs with another column                             \\
7           & Swap the $layer\_to\_chip$ entries of one batch with another batch                                            
\end{tblr}
\end{table}

The mapping encoding scheme introduced in Section \ref{sec:mapping encoding scheme} defines the mapping and scheduling space of Compass. The mapping generation engine is responsible for efficiently exploring and sampling within this space. We adopt a genetic algorithm (GA) for exploration. This engine explores values for $segmentation$ and $layer\_to\_chip$, while $micro\_batch\_size$ is handled by the hardware sampling engine. This separation is due to the high overhead associated with updating $micro\_batch\_size$, which requires re-fusing layer workloads and regenerating the model, making it unsuitable for rapid exploration in this stage.

\textbf{Selection Operation}: Tournament selection is used to filter offspring from the parent population. This method randomly selects individuals from the parents with equal probability, and then chooses the one with the highest fitness to be part of the offspring. Tournament selection does not tie selection probability strictly to fitness, can adapt to multiple optimization objectives, and simultaneously avoids population degradation.

\textbf{Crossover Operation}: Bitwise crossover is applied to $segmentation$. Each bit is randomly inherited from one of the two parent individuals. For $layer\_to\_chip$, a subgraph-level crossover is used. Subgraphs are first determined based on the crossover result of $segmentation$, and then one parent is randomly selected to contribute its corresponding subgraph’s $layer\_to\_chip$ values to the offspring. This approach balances randomness and local stability of the computation graph.

\textbf{Mutation Operation}: For $segmentation$, two mutation operators are defined: bit-flip and bit-swap. Bit-flip randomly selects a bit and toggles its value between 0 and 1. Bit-swap randomly selects a position and swaps its value with either the previous or the next bit. For $layer\_to\_chip$, we define seven mutation operators (see Table \ref{tab:mutation_types}), ensuring complete coverage of the mapping space. Operators 1–3 have minimal impact, modifying only a single layer; operators 4–5 have moderate impact at the subgraph level; and operators 6–7 introduce broader changes by modifying entire rows or columns of the computation graph. During runtime, mutation probability is dynamically adjusted based on the exploration phase: more impactful mutations are favored in early stages for broader exploration, while less impactful ones are prioritized later for fine-tuning convergence.

\subsection{Hardware Sampling Engine}

\subsubsection{\textbf{Problem Definition}}

This section introduces the hardware sampling engine constructed for the design of heterogeneous multi-chiplet accelerators. In real-world scenarios, customizing distinct chiplets from scratch for every specific workload incurs exorbitant costs and long development cycles. The prevailing industry trend is to integrate and reuse pre-fabricated or pre-designed chiplets via advanced packaging technologies. Based on this, we formulate the heterogeneous multi-chiplet architecture search problem as follows:

Assume the existence of a pre-built heterogeneous chiplet library $\mathcal{T}$. The chiplets in this library provide design options across two orthogonal dimensions: 1) \textbf{Computing capacity}: defines the computational resource volume of the chiplet (e.g., 1K MACs or 4K MACs); 2) \textbf{Dataflow type}: defines the internal microarchitecture for data reuse within the chiplet (e.g., weight stationary or output stationary).

Given the overall computational performance requirement of the target accelerator, our objective is to sample and construct the optimal hardware parameter configuration from a discrete combinatorial space. This process can be decoupled into the following sequential decision paths:

\begin{itemize*}[itemjoin=\\\hspace*{\parindent}]
        \item Determining the array dimension ($z_{shape}$): The engine first selects a uniform computing capacity from the library. Since the total system computational target is a given hard constraint, the selected single-chiplet specification inversely dictates the total number of chiplets $N$ required by the system. Consequently, the engine determines the interconnect array dimension $z_{shape} = (H, W)$ on the package substrate.
    \item Configuring the chiplet layout ($z_{layout}$): Once the array dimension is fixed, each slot within the array can independently be assigned a different dataflow type from the chiplet library. The topological arrangement of these heterogeneous chiplets across the 2D space constitutes the chiplet layout $z_{layout} \in \mathcal{T}_{types}^{H \times W}$.
    \item Determining global system parameters ($z_{sys}$): In addition to the internal topological structure of the chiplet array, the engine must jointly optimize global system parameters shared across the entire accelerator, such as NoP bandwidth and DRAM bandwidth.
\end{itemize*}

Synthesizing the decisions above, our goal is to identify the optimal joint configuration tensor $Z = [z_{sys}, z_{shape}, z_{layout}]$ that minimizes the comprehensive cost under the target workload. Because every generated hardware architecture must be evaluated via the mapping search process, the time cost for a single architecture evaluation is relatively high. To guarantee highly efficient and valuable architecture sampling, this engine employs an enhanced Bayesian optimization (BO) method.

\subsubsection{\textbf{Bayesian Optimization Method}}

A Gaussian process (GP) is employed as the surrogate model to achieve smooth fitting of the objective function using a limited number of samples. Expected improvement (EI) is utilized as the acquisition function to guide the subsequent round of architecture sampling, effectively balancing exploitation and exploration. The GP quantifies the prior similarity between different hardware configurations $Z$ and $Z'$ via a covariance kernel function. Since our design space inherently involves a 2D chiplet layout structure, applying a traditional radial basis function (RBF) kernel directly to a flattened 1D vector would obliterate the spatial topological relationships among the chiplets. To address this, we design a hardware-aware composite kernel function:

\begin{equation}
\begin{split}
K(Z, Z') = K_{sys}(z_{sys}, z_{sys}') \cdot \Big[ 1 + \mathbb{I}(z_{shape} = z_{shape}') \\ \cdot
K_{layout}(z_{layout}, z_{layout}') \Big]
\end{split}
\end{equation}

Here, $K_{sys}$ targets system-level parameters such as NoP bandwidth. Because hardware parameters are typically selected from a predefined set of discrete candidates, we employ a standard RBF kernel in a continuous space for distance measurement, and map the results to discrete configuration indices via nearest-integer rounding. $\mathbb{I}(\cdot)$ is an indicator function, which ensures that the internal layout similarity is evaluated if and only if the array dimensions of the two configurations are identical. $K_{layout}$ denotes the chiplet layout kernel, which cross-compares all slots between two layout grids: when the chiplet types deployed in two slots are identical, it contributes to the similarity score based on their topological proximity. Let $z_{layout}[u]$ denote the chiplet dataflow type deployed at the 2D coordinate $u$ in layout $z_{layout}$. The formal definition of $K_{layout}$ is as follows:

\begin{equation}
\begin{split}
K_{layout}(z_{layout}, z_{layout}') = \sigma_{layout}^2 \sum_{u} \sum_{v} \\
\mathbb{I}\big(z_{layout}[u] = z'_{layout}[v]\big) \cdot \mathbf{W}_{u,v}
\end{split}
\end{equation}

Where $u$ and $v$ iterate over all 2D coordinates within the layout grid. $\mathbf{W}$ is the positional similarity weight matrix, which decays exponentially with the Manhattan distance between the two coordinates. This implicitly reflects the routing hop counts and communication costs among chiplets:

\begin{equation}
\begin{split}
    \mathbf{W}_{u,v} = \exp\left(-\frac{|x_u - x_v| + |y_u - y_v|}{\lambda_{layout}}\right)
\end{split}
\end{equation}

The variance $\sigma_{layout}^2$ and the length scale $\lambda_{layout}$ in the formulas are hyperparameters of the model, which are learned automatically during the BO. 

Finally, because the input configurations consist entirely of discrete variables, it is unfeasible to directly use gradient-based methods to find the next candidate solution that maximizes the EI. Therefore, we design a two-tier simulated annealing algorithm to conduct sampling exploration on the surrogate model. The outer-tier mutation operator is responsible for macroscopic perturbation by randomly selecting a dimension from $z_{shape}$ and $z_{sys}$. The inner tier executes fine-grained adjustments specifically on $z_{layout}$. If the array dimension changes, it triggers a reallocation mapping of the inner layout; if the dimension remains unchanged, it executes either a single-slot random replacement or a dual-slot chiplet position swap to explore the optimal chiplet layout.

\subsection{Evaluation Engine}

The evaluation engine models latency, energy, and monetary cost for a given mapping and hardware configuration by combining intra-chiplet and inter-chiplet evaluations.

\textbf{Intra-Chiplet Evaluation}: We employ ZigZag \cite{zigzag} to obtain the energy and latency for individual layers, and perform operation counting for non-compute-intensive layers through the post-processing unit, thereby enabling evaluation of layers such as activation and normalization. In addition, we extend this foundation with support for temporal and spatial tiling by computing data reuse factors under different tiling strategies. Spatial tiling is similar to tensor parallelism \cite{tensor-parall} or spatial mapping \cite{gemini} in prior works, where a layer is divided into multiple sub-layers to be processed by different chiplets. Temporal tiling, on the other hand, addresses insufficient buffer capacity by partitioning the workload into sub-blocks based on the maximum available buffer size.

\textbf{Inter-Chiplet Latency}: The total processing latency for a single layer is determined by the maximum of its computation latency and data access latency from DRAM and NoP, based on the widely used double-buffering assumption: $ T_{\text{proc}, l} = \max \left( T_{\text{comp}, l}, T_{\text{DRAM}, l}, T_{\text{NoP}, l} \right)$. The evaluation engine simulates layer execution based on the scheduling order defined in the mapping encoding. The start time of each layer is determined by the later of two times: the completion time of the previously scheduled layer on the same core, and the latest completion time among all its direct predecessor layers: $
T_{\text{start}, l} = \max\left( 
    \max_{l' \in \text{Pre}(l)} T_{\text{end}, l'},\ 
    \max_{l'' \in \text{SameCore}(l)} T_{\text{end}, l''}
\right)
$. The completion time for a layer is then: $T_{\text{end}, l} = T_{\text{start}, l}+T_{\text{proc}, l}$. The model's total execution time is the maximum completion time across all layers: $T_{\text{model}} = \max_{l \in \mathcal{L}} T_{\text{end}, l}$.

\textbf{Inter-Chiplet Energy}: The total processing energy of a single layer is the sum of compute energy and data access energy from DRAM and NoP: $ E_{\text{proc}, l} =  E_{\text{comp}, l}+E_{\text{DRAM}, l}+E_{\text{NoP}, l}$. The total energy for the entire model is the sum of energy consumed by all layers: $E_{\text{model}} = \sum_{l \in \mathcal{L}} T_{\text{proc}, l}$.

\begin{algorithm}[tb]
\caption{Data Access Flag Determination}
\small
\label{alg:data access flag}
\KwIn{$\textit{}$
Neural network model: $\textit{model}$; computation execution graph: $\textit{execGraph}$ and corresponding data dimensions: $\textit{segDim}$, $\textit{batchDim}$, $\textit{subGraphSize}$; number of chiplets: $\textit{chipNum}$;
}

\KwOut{
Sets of direct predecessor and successor layers for each layer: $\textit{layersPrev}$, $\textit{layersNext}$; data access flags for each layer: $\textit{isWriteOut}$, $\textit{isLoadWei}$;

}
\KwData{
    Chiplet status table: \textit{chipState};
}

Initialize $\textit{isWriteOut}$ with $\textbf{true}$\; Initialize $\textit{isLoadWei}$ with $\textbf{true}$\;
$(\textit{layersPrev}, \textit{layersNext}) \gets$ \texttt{parse}($\textit{model}$)

\For{$i \gets 0$ \KwTo segDim}{
    \For{$j \gets 0$ \KwTo batchDim}{
        \For{$k \gets 0$ \KwTo subGraphSize}{
            $(\textit{currChip}, \textit{currLayer}) \gets \textit{execGraph}[i, j, k]$\;
            \For{$c \gets 0$ \KwTo chipNum}{
                $(\textit{prevBatch}, \textit{prevLayer}) \gets \textit{chipState}[c]$\;
                
                \If{$c = \textit{currChip} \land \textit{prevLayer} = \textit{currLayer} \land j \ne \textit{prevBatch}$}{
                    $\textit{isLoadWei}[j][\textit{currLayer}] \gets \textbf{false}$\;
                }
                
                \If{prevBatch = $j$}{
                    $\textit{layersNext}[\textit{prevBatch}][\textit{prevLayer}]$
                    $.\texttt{erase}$$(\textit{currLayer})$\;
                    
                    \If{$\textit{layersNext}[\textit{prevBatch}][\textit{prevLayer}]$
                    $.\texttt{empty()}$}{
                        $\textit{isWriteOut}[\textit{prevBatch}][\textit{prevLayer}]$ $\gets$ $\textbf{false}$\;
                    }
                    
                    $\textit{layersPrev}[$j$][\textit{currLayer}]$
                    .\texttt{erase}$(\textit{prevLayer})$\;
                }
            }
            $\textit{chipState}[\textit{currChip}] \gets (j, \textit{currLayer})$\;
        }
    }
}
\Return $\textit{isWriteOut}$, $\textit{isLoadWei}$, $\textit{layersPrev}$, $\textit{layersNext}$\;

\end{algorithm}

\textbf{Data Access Analysis}: As seen above, determining DRAM and NoP data access patterns is key to modeling inter-chiplet latency and energy. We use a scanning algorithm (see Algorithm \ref{alg:data access flag}) for this purpose. The algorithm maintains a chiplet status table and sets of direct predecessor and successor layers for each layer. The status table tracks which layers are temporarily stored on each chiplet during traversal. The basic idea is: if a layer is evicted from a chiplet and all its successor layers have already appeared (and it’s not the last layer), then its output does not need to be written back to off-chip memory. When a layer is executed and the previous layer executed on the same chiplet has the same layer index but a different micro-batch ID, it indicates that weights do not need to be reloaded. The source of activations from predecessor layers is determined using the layersPrev set maintained by the algorithm. If a predecessor layer is still in layersPrev, it implies the output must be fetched from off-chip memory; otherwise, it can be directly retrieved from another chiplet via NoP. Additionally, Compass supports setting mandatory result write-out flags on a per-layer basis and allows specifying the DRAM ID for each layer's off-chip input and output. This enables flexible memory management control to accommodate the KV Cache management requirements in LLMs.

\textbf{Monetary Cost}: Compass evaluates monetary cost based on area and bandwidth. We use the yield formula from Gemini\cite{gemini}: $Y_c = Y_{\text{unit}}^{\frac{A_c}{A_{\text{unit}}}}$, where $Y_\text{unit}$ and $A_\text{unit}$ represent standard yield and area, and $Y_c$ and $A_c$ represent yield and area for the given chiplet. NoP and DRAM contribute to the overall cost by increasing the area of chiplets and IO dies through their bandwidth. The area of a single chiplet $c$ is: $A_c=A_{\text{MAC},c}+A_{\text{SRAM},c}+A_{\text{NoC},c}+\alpha \cdot BW_{\text{NoP}}+A_{\text{Others}}$. The cost of a single chiplet is: $MC_{c}=A_c/Y_c\cdot COST_{\text{chip}}$. The area of a IO die is: $A_{\text{IO}}=\beta \cdot BW_{\text{NoP}}+\gamma \cdot BW_{\text{DRAM}}$. The cost of a IO die is: $MC_{\text{IO}}=A_{\text{IO}}/Y_{\text{IO}}\cdot COST_{\text{IO}}$.
The package area is scaled proportionally based on the total area of all chiplets, yielding a package cost represented as: $MC_{\text{pack}}=(\sum_{c \in C} A_c+\sum A_{IO})\cdot COST_{\text{pack}}$. The total monetary cost is: $MC_{\text{total}} = \sum_{c \in C} MC_{c}+ \sum MC_{IO}+ MC_{\text{pack}}$.

\section{Experiments and Evaluation}\label{sec:exp}

\subsection{Experimental Setup}

\begin{table}
\centering
\caption{The range of hardware parameters for DSE}
\label{table:exp:hardware range}
\begin{tblr}{
  width = \linewidth,
  colspec = {Q[100]Q[450]Q[450]},
  row{4} = {c},
  row{5} = {c},
  row{6} = {c},
  row{7} = {c},
  columns = {m},
  cell{1}{2} = {c},
  cell{1}{3} = {c},
  cell{2}{2} = {c},
  cell{2}{3} = {c},
  cell{3}{1} = {r=5}{},
  cell{3}{2} = {c},
  cell{3}{3} = {c},
  hlines,
  vlines,
}
$z_{shape}$ & Chiplet Spec (MACs, GLB)   & S (1K, 2MB), M (4K, 8MB), L (16K, 32MB)    \\
$z_{layout}$ & Chiplet Dataflow           & \textbf{}Weight Stationary (WS), Output Stationary (OS) \\
$z_{sys}$ & NoP Bandwidth              & {[}32, 64, 128, 256, 512] GB/s                          \\
   & Bandwidth per DRAM Chip    & {[}16, 32, 64, 128, 256] GB/s                           \\
   & Micro-Batch Size (Prefill) & {[}1, 2, 4]                                             \\
   & Micro-Batch Size (Decode)  & {[}1, 2, 4, 8, 16, 32, 64, 128]                         \\
   & Tensor Parallelism         & {[}8, 16, 32, 64] (number of FFN layer partitions)      
\end{tblr}
\end{table}

\textbf{DSE Setup.} In this work, we design DNN accelerators with compute capacities of 64 TOPS, 512 TOPS, and 2048 TOPS to demonstrate considerations for using chiplet technology in different scenarios. The 64 TOPS is comparable to that of Simba\cite{simba}, representing edge computing scenarios. The 512 TOPS and 2048 TOPS configurations correspond to the compute capacities of the A100\cite{a100} and H100\cite{h100} GPUs, representing medium- and large-scale scenarios. The design objective is to minimize the product of latency, energy, and monetary cost.

\textbf{Workload Setup.} We target GPT3-7B, GPT3-13B\cite{gpt3}, and LLaMA3-70B\cite{llama,llama3} as the LLMs for the 64 TOPS, 512 TOPS, and 2048 TOPS configurations, respectively. GPT3 adopts a traditional transformer architecture, while LLaMA3 features SOTA techniques like GQA\cite{GQA} and pre-layer normalization\cite{pre_layer_norm}. The parameter scales of the models are aligned with the corresponding compute capacities. For each configuration, we perform DSE separately for prefill and decode tasks. The batch sizes for the prefill and decode phases are set to 4 and 128, respectively. These values are empirically chosen based on prevalent configurations in state-of-the-art benchmarks (e.g., MLPerf\cite{mlperf}) and inference serving engines (e.g., vLLM\cite{vllm}).

\begin{figure*}[ht]
\centering
\includegraphics[width=0.99\textwidth]{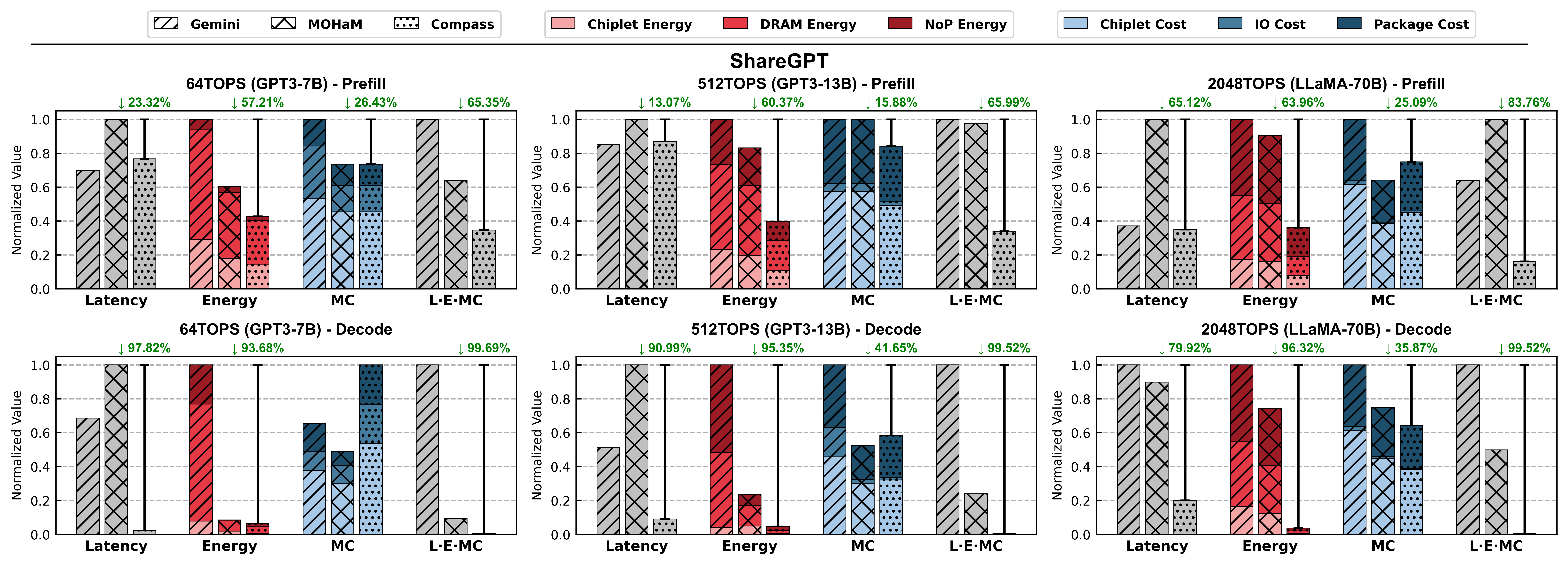}
\includegraphics[width=0.99\textwidth]{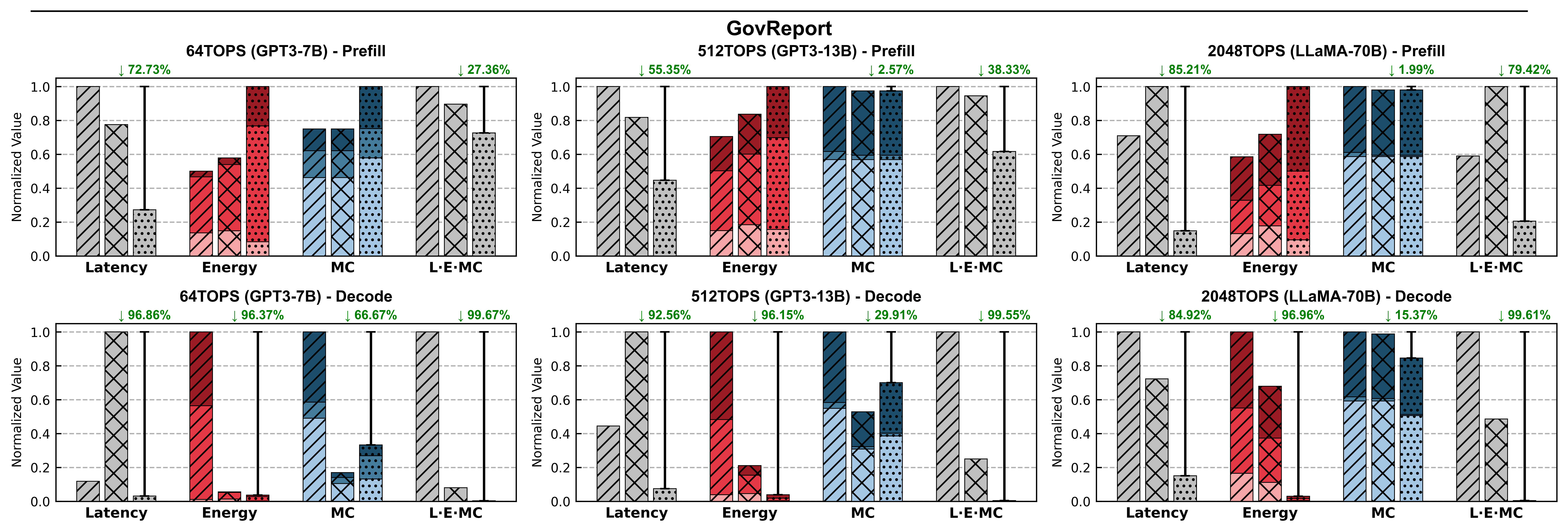}
\caption{Comparison results among Gemini, MOHaM, and Compass. Each metric is normalized by setting the maximum value within its comparison group to 1. \textbf{The hatch patterns in the bars indicate different baselines, while the colors represent the breakdown of a specific metric}.}
\label{fig:exp_diff}
\end{figure*}

\textbf{Scenario Setup.} Considering typical LLM application scenarios, we use the ShareGPT (dialogue)\cite{sharegpt} and GovReport (summarization)\cite{govreport} datasets for DSE. ShareGPT has average input and output sequence lengths of 78 and 483, exhibiting short-input, long-output characteristics. GovReport has averages of 9652 and 602, showing long-input, short-output characteristics. We randomly sample from each dataset to build a fitting set and a test set. The fitting set simulates historical sequence length distributions to guide DSE, while the test set simulates unseen tasks to validate and compare DSE results.

\textbf{Baseline Setup.} Gemini\cite{gemini} and MOHaM\cite{moham} are selected as comparison baselines. Both methods are adapted to convert into the mapping method of Compass. Gemini, as a single-model DSE work, only supports DSE for fixed sequence lengths; therefore, we perform DSE with the average sequence length of the scenario to achieve a relatively fair comparison. For mapping, it uses the simulated annealing method. For hardware, it only supports arrays with homogeneous dataflow and uses grid search. MOHaM is a multi-model DSE work that treats each micro batch as a different model to partially support dynamicity. It uses a genetic algorithm to jointly optimize mapping and hardware configurations.

\textbf{Hardware Parameters.} We adopt the same DSE-independent hardware parameters as Gemini: TSMC 12 nm process technology, organic substrate packaging, 1 GHz clock frequency, NoP employing GRS technology, based on a mesh topology with XY routing algorithm, and 4 DRAM chips evenly distributed on the left and right sides of the chiplet array. Table \ref{table:exp:hardware range} lists the candidate values of the hardware parameters relevant to DSE. Changes in micro batch size and tensor parallelism cause the entire computational execution graph to change, therefore they are also searched as hardware system parameters.

\textbf{Hyperparameters and Environment.} The GA population size is set to 120 with multithreaded parallel evaluation and 100 iterations. The number of iterations for Bayesian optimization is 100. The server used for the experiments has two Hygon 7285 processors with 128 logical cores, and is equipped with an A100 GPU to update the Bayesian model parameters. Under this configuration, the mapping search and Bayesian optimization update for a single hardware configuration take an average of 3 minutes, and the hardware configuration search under a single workload scenario takes an average of about 5 hours, which is acceptable for the actual hardware design process.

\subsection{Validation Experiments}

\begin{table}
\centering
\caption{Verification comparison with Gemini}
\label{tab:exp_val}
\begin{tblr}{
  colspec = {cccccc},       
  cells = {c, m},          
  hlines,                  
  vlines,                  
  row{1,2} = {font=\bfseries}, 
}
\SetCell[r=2]{c} & \SetCell[r=2]{c} {MC ($\$$)} & \SetCell[c=2]{c} Prefill & & \SetCell[c=2]{c} Decode & \\
 & & L ($10^6$) & E ($10^{10}$) & L ($10^6$) & E ($10^{10}$) \\
\textbf{Gemini} & 2424.6 & 210.76 & 53.48 & 3.41 & 8.23 \\
\textbf{Compass} & 2424.6 & 210.76 & 54.52 & 3.41 & 8.01 \\
\textit{Error}& 0.00\% & 0.00\% & 1.91\% & 0.00\% & 2.71\% \\
\end{tblr}
\end{table}

Gemini is a simulator whose accuracy has been verified by real-world chips. Therefore, we conduct a calibration comparison between Compass and Gemini to demonstrate the effectiveness of the Compass evaluation engine. We provide Compass and Gemini with the same Simba-like hardware configuration, adopting the same layer-pipeline mapping strategy. The workload is based on the GPT3-7B model, using the same batch size and sequence length workloads for verification. Table \ref{tab:exp_val} shows the verification results. It can be seen that compared to Gemini, the errors of the evaluation engine designed by Compass in latency (L), energy (E), and monetary cost (MC) are all within 3\%, demonstrating the accuracy of the Compass.

\subsection{Comparative Experiments}\label{sec:exp diff}

Fig. \ref{fig:exp_diff} shows the comparison of latency, energy, monetary cost, and total cost of the solutions found by the three methods under various scenario combinations. It can be seen that Compass finds better solutions under all scenarios compared to the other two methods. Compared to Gemini, Compass reduces latency, energy, monetary cost, and total cost by an average of 58.5\%, 45.33\%, 14.57\%, and 77.87\%, respectively; compared to MOHaM, with a 20.8\% increase in monetary cost, it achieves an average reduction of 63.92\%, 40.32\%, and 77.05\% in latency, energy, and total cost.

Compared to the prefill phase, the solutions provided by Compass under the decode phase can reduce latency and energy to a greater extent. This is mainly because the decode phase has a larger batch size. As described in Section \ref{sec:intro}, the dynamicity of LLM inference serving is mainly reflected in the different request types and sequence lengths within a batch; therefore, a larger batch size leads to stronger dynamicity. Gemini uses a fixed sequence length scheme, resulting in redundant padding; while MOHaM treats each micro batch as a separate model, preventing the possibility of merging them for batch processing in the QKV generation and FFN stages. Both schemes lead to an incomplete mapping space for dynamic LLM workloads, resulting in worse latency and energy performance.

In addition, Compass also demonstrates the flexibility of DSE under different scenarios. For example, in the GovReport-Prefill scenario, the total costs of the solutions provided by the three methods are similar. Compass reduces more latency by slightly sacrificing energy and monetary cost, thereby achieving a lower total cost.

\subsection{Hardware Design Analysis}

\begin{table*}
\centering
\caption{The optimal hardware architecture configurations searched by Compass}
\label{exp:solution}
\scalebox{0.9}{
\begin{tblr}{
  width = \linewidth,
  colspec = {Q[235]Q[42]Q[58]Q[73]Q[42]Q[58]Q[73]Q[42]Q[58]Q[73]Q[42]Q[58]Q[73]},
  cells = {c},
  cell{1}{1} = {r=3}{},
  cell{1}{2} = {c=6}{0.346\linewidth},
  cell{1}{8} = {c=6}{0.346\linewidth},
  cell{2}{2} = {c=3}{0.173\linewidth},
  cell{2}{5} = {c=3}{0.173\linewidth},
  cell{2}{8} = {c=3}{0.173\linewidth},
  cell{2}{11} = {c=3}{0.173\linewidth},
  hlines,
  vlines,
}
                   & ShareGPT &     &      &        &     &      & GovReport &     &      &        &     &      \\
                   & Prefill  &     &      & Decode &     &      & Prefill   &     &      & Decode &     &      \\
                   & 64       & 512 & 2048 & 64     & 512 & 2048 & 64        & 512 & 2048 & 64     & 512 & 2048 \\
DRAM\_BW           & 16       & 16  & 32   & 32     & 16  & 16   & 16        & 16  & 16   & 64     & 16  & 16   \\
NoP\_BW            & 32       & 32  & 64   & 64     & 32  & 32   & 32        & 32  & 32   & 32     & 32  & 32   \\
Micro\_batch\_size & 4        & 4   & 4    & 64     & 128 & 64   & 1         & 2   & 4    & 64     & 128 & 64   \\
Tensor\_Parall     & 4        & 16  & 32   & 16     & 32  & 32   & 64        & 32  & 64   & 4      & 32  & 32   \\
Chiplet Spec       & L        & L   & L    & M      & M   & L    & M         & L   & L    & M      & M   & L    \\
WS Number             & 2        & 13  & 38   & 8      & 30  & 34   & 0         & 0   & 10   & 8      & 30  & 34   \\
OS Number             & 0        & 3   & 26   & 0      & 34  & 30   & 8         & 16  & 54   & 0      & 34  & 30   
\end{tblr}
}
\end{table*}

Table \ref{exp:solution} shows the optimal hardware configurations searched by Compass. We conduct a comparative analysis of these configuration parameters combined with the scenarios.

\textbf{Bandwidth.} In the prefill phase, when the compute scale is small, ShareGPT is compute-bound, so it maintains a lower bandwidth. After the compute scale reaches 2048, bandwidth becomes the bottleneck and increases. However, due to long sequence lengths, GovReport is compute-bound across all compute scales, consistently maintaining a lower bandwidth demand. In the decode phase, when the model is small, the bottleneck lies in the MHA stage performing GEMV operations with low compute intensity, requiring larger bandwidth. As the model parameters increase with the compute scale, the bottleneck shifts to the GEMM in the FFN stage, becoming compute-bound, and thus the bandwidth demand decreases.

\textbf{Micro Batch Size and Parallelism.} The micro batch size in ShareGPT-prefill always maintains the maximum value due to short sequence lengths and insufficient compute intensity. In GovReport-prefill, it gradually increases with the compute scale due to longer input sequences. In the decode phase, it maintains high values (64 and 128) due to lower compute intensity. Tensor parallelism is responsible for partitioning the FFN computation. It can be seen that it generally increases with the increase of model scale and compute scale to fully utilize the parallelism of the chiplets.

\textbf{Chiplet Specifications and Types.} In the selection of chiplet specifications, small-spec chiplets are not selected because this increases the number of chiplets and mapping difficulty, which leads to low utilization due to the task dependencies processed on the chiplets. The prefill phase prefers large-spec chiplets, while the decode prefers medium-spec chiplets. This is because in the prefill phase, using a single chiplet with larger compute capacity to process a single large-scale matrix operation is more cost-effective. In the decode phase, due to a larger batch size, it prefers to appropriately increase the number of chiplets for parallel processing.

\textbf{Sequence Length Distribution.} Under the two sequence length distributions, the architectures found in the decode phase are similar; especially in chiplet specifications and chiplet layout, they are identical. This may be attributed to their similar output sequence lengths. However, it is different in the prefill phase. On the Sharegpt side, the number of WS chiplets is the majority. This is because under this distribution, the sequence length is short, the model parameters dominate, and the weight matrix is larger, making it more suitable to be paralleled and fixed. Under GovReport, the sequence is long, and the input and output matrices in GEMM are larger, so it prefers output stationary chiplets.

The aforementioned hardware architecture parameter distribution reveals the complexity of large model inference system design. Not only does the expansion of compute scale change the bottleneck, but the fluctuation of sequence length and the switching of execution phases also change the accelerator's demand for memory access, communication, and parallelism. Therefore, the design of multi-chiplet accelerators should abandon the one-size-fits-all approach and conduct separate designs under different workloads and scenarios. This also demonstrates the necessity of Compass, which can provide automated and efficient design exploration for scenario-driven accelerator design.

\subsection{Mapping Design Analysis}

\begin{figure}[ht]
\centering
\includegraphics[width=0.49\textwidth]{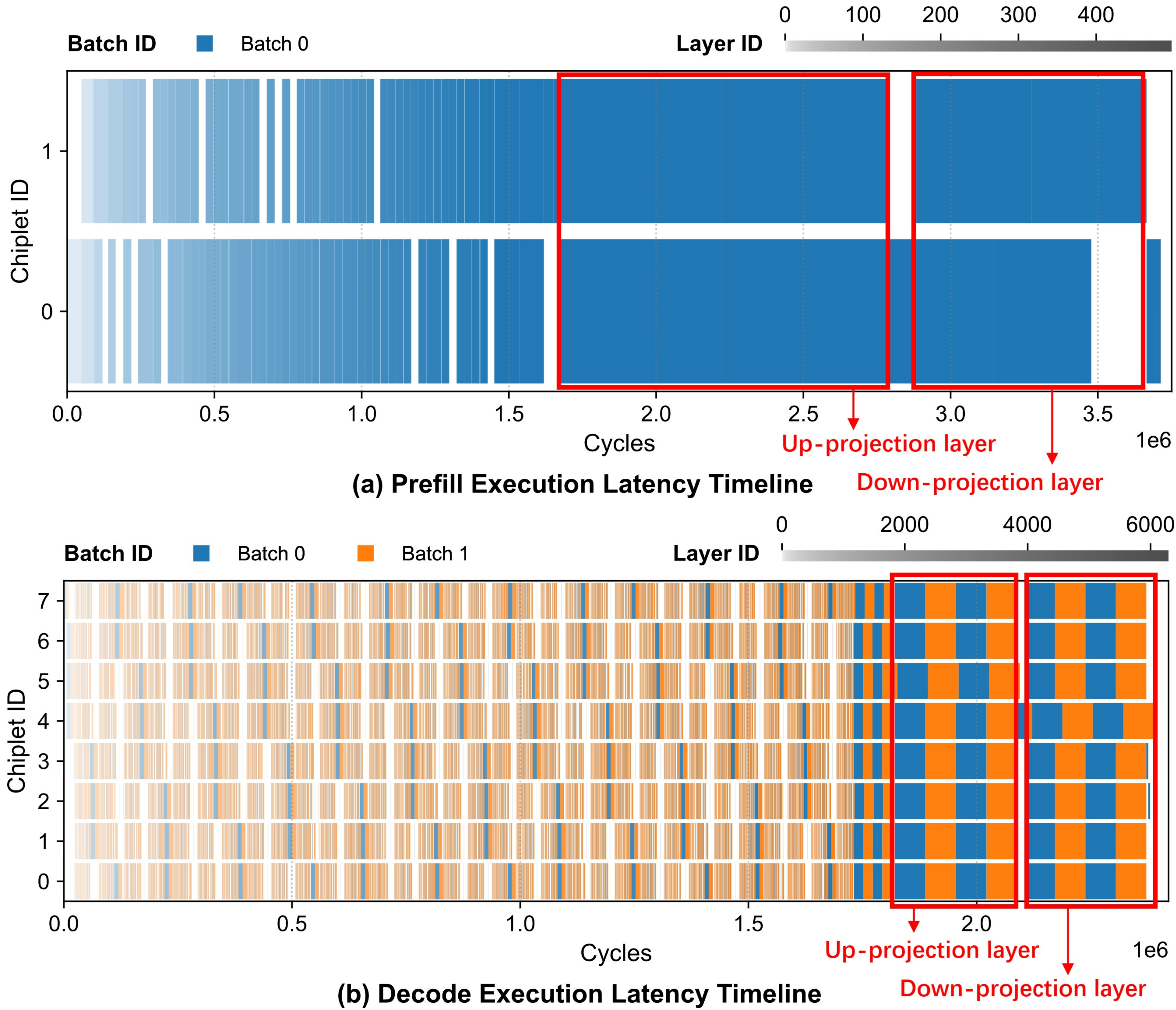}
\caption{Execution latency timeline of a single LLM block under ShareGPT-64TOPS.}
\label{fig:exp:timeline}
\end{figure}

We take the mapping scheme searched under ShareGPT-64TOPS as a case for analysis. Fig. \ref{fig:exp:timeline} shows the execution latency timeline under this scenario.

During the prefill phase, as indicated by the searched system parameters in Table \ref{exp:solution}, the optimal micro-batch size is 4 (matching the total batch size). This is because shorter sequence lengths necessitate assembling larger micro-batches to increase computation intensity. At this time, the entire model does not distinguish batches, exhibiting an execution manner similar to model parallelism, and multiple layers are distributed across chiplets as much as possible for parallel execution.

Under the decode scenario, two micro-batches are constructed, and the tensor parallelism degree is set to 16, meaning the up-projection and down-projection layers in the FFN are both tiled into 16 parts for parallelism. At this time, the execution of the entire model presents an execution manner similar to pipeline parallelism. According to Fig. \ref{fig:exp:timeline}(b), from the mapping of the up and down projections, it can be seen that the sixteen tiled sub-layers are divided into two groups and executed on 8 chiplets. Each group switches to the next group only after all batches have been executed, enabling the weights to reside on the chiplets for reuse, thereby reducing the cost of data movement.

From the two cases above, it can be seen that Compass can flexibly use different mapping strategies for different workloads to optimize model execution.

\subsection{Integration with SOTA Inference Service Scheduling}\label{case study:hybrid sched}

Section \ref{sec:exp diff} compares prefill and decode workloads separately, in order to enable a fair comparison with prior work. However, SOTA LLM inference servers adopt more advanced scheduling strategies and often orchestrate prefill and decode requests together. This behavior is not supported in prior works such as Gemini. This section explores the interplay between these strategies and multi-chiplet accelerators, thereby highlighting the importance of mixed request orchestration and heterogeneous design.

We take the GovReport-512TOPS scenario as a case study for analysis, because this configuration targets the more common long-context scenarios and considers the compute demands of both edge and cloud. In this scenario, the average output sequence length is 602, meaning the ratio of prefill requests to decode requests is approximately 1:602. Therefore, the DSE workload can be defined as jointly optimizing 1 group of prefill requests with a batch size of 1 and 5 groups of decode requests with a batch size of 128.

\begin{figure}[ht]
\centering
\includegraphics[width=0.43\textwidth]{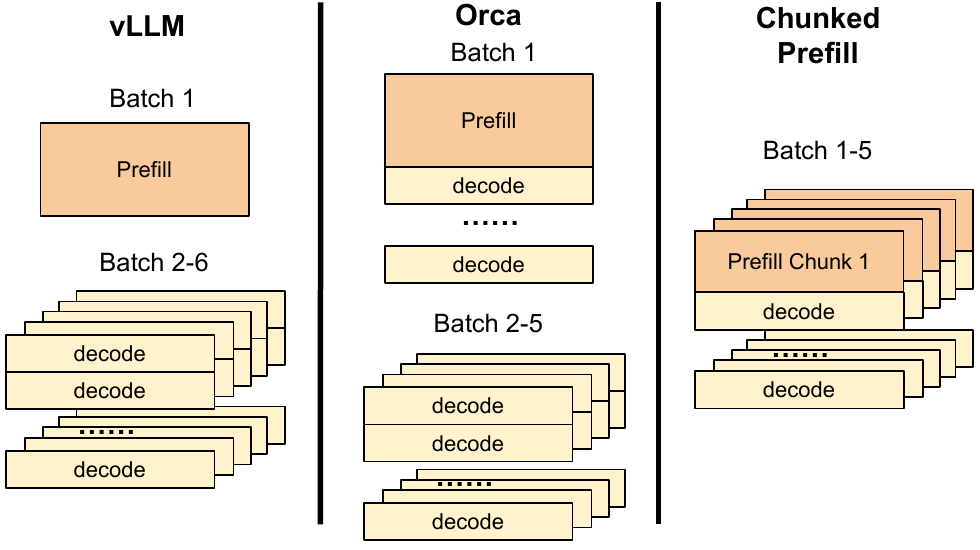}
\caption{Three SOTA inference service scheduling strategies.}
\label{fig:case study:hybrid req}
\end{figure}

\begin{figure}[ht]
\centering
\includegraphics[width=0.49\textwidth]{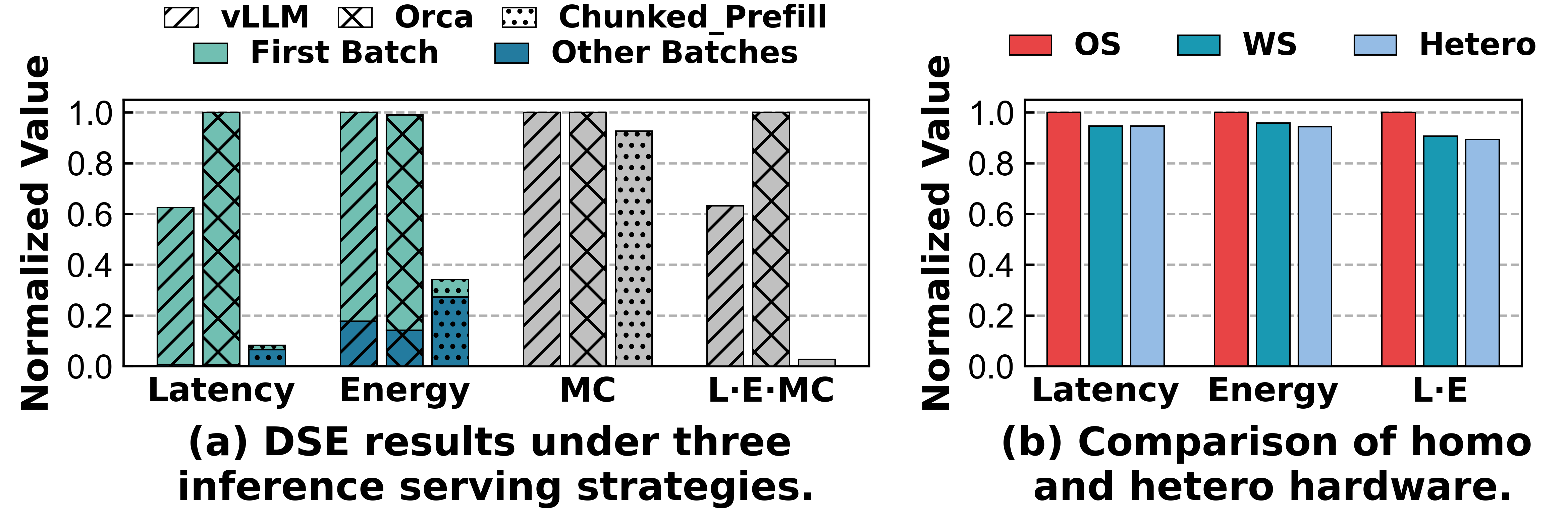}
\caption{DSE results for different serving strategies and comparison of homogeneous and heterogeneous hardware.}
\label{fig:case study:hybrid diff}
\end{figure}

Fig. \ref{fig:case study:hybrid req} illustrates three SOTA workload orchestration strategies. vLLM adopts a separated strategy, treating the prefill request as a standalone batch. Orca uses a mixed strategy, orchestrating the prefill request together with other decode requests. Chunked Prefill splits the entire prefill process into several chunks, dispatching each chunk into different batches to be processed alongside other decode requests.

\begin{table}
\centering
\caption{Optimal hardware parameters found under three inference serving strategies}
\label{table: hybrid_hardware}
\scalebox{0.9}{
\begin{tblr}{
  width = \linewidth,
  colspec = {Q[148]Q[183]Q[215]Q[140]Q[98]Q[104]},
  cells = {c},
  hlines,
  vlines,
}
      & DR BW & NoP BW & Spec & WS & OS~ \\
vLLM  & 64    & 32     & L    & 3  & 13  \\
Orca  & 64    & 32     & L    & 3  & 13  \\
C. P. & 16    & 32     & L    & 12 & 4   
\end{tblr}
}
\end{table}

We use Compass to explore the optimal hardware mapping schemes under three strategies. Fig. \ref{fig:case study:hybrid diff}(a) shows the latency, energy, and MC of these schemes, and provides the latency and energy breakdown of each scheme when processing the first batch and other batches. Table \ref{table: hybrid_hardware} presents the hardware architecture parameters of the three schemes.

It can be seen that for vLLM and Orca, the first batch contributes most of the latency and energy, which mainly originates from the longer input sequence length caused by GovReport. This further affects hardware selection, making the hardware parameters found by vLLM and Orca basically identical, with slight differences only in the specific chiplet layout. Because the Chunked Prefill strategy processes the entire prefill request in chunks, it makes each batch have similar latency and energy. From the final latency and energy in Fig. \ref{fig:case study:hybrid diff}(a), processing long-sequence prefill requests in chunks can balance the utilization of bandwidth and computation, effectively reducing the overall cost, which also aligns with real-world practice. From the chiplet types in Table \ref{table: hybrid_hardware}, vLLM and Orca are dominated by the prefill of long sequence lengths, thus using more OS chiplets; Chunked Prefill is closer to the decode computation pattern, and the shorter sequence length makes it more advantageous to use WS chiplets to parallelize the weight matrix.

Further, we replace all chiplet types of the hardware architecture found under the chunked prefill strategy with OS and WS to compare the differences between homogeneous and heterogeneous architectures. Fig. \ref{fig:case study:hybrid diff}(b) shows the comparison results. It can be seen that the heterogeneous architecture has the minimum cost, demonstrating the advantage of heterogeneous design. Compared to the standalone OS and WS architectures, the EDP is reduced by 10.7\% and 1.5\%, respectively. In addition, the standalone WS architecture is better than the standalone OS architecture and is close to the heterogeneous architecture, which matches the phenomenon that there are more WS-type chiplets in the heterogeneous architecture.

In summary, the choice of serving strategy leads to workload changes, further affecting hardware selection. Therefore, it is necessary to conduct design exploration combined with the applicable sequence length distribution and serving strategy.

\subsection{Ablation Study}

\begin{figure}[ht]
\centering
\includegraphics[width=0.49\textwidth]{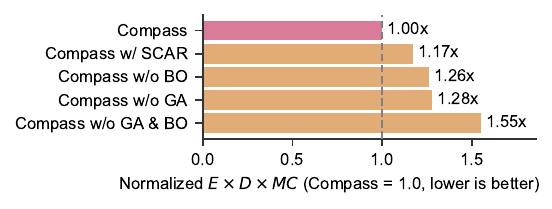}
\caption{Comparison results after ablating each component in Compass.}
\label{fig:ablation}
\end{figure}

We further construct ablation experiments under the chunked prefill configuration in Section \ref{case study:hybrid sched} to demonstrate the effectiveness of the mapping generation engine and the hardware sampling engine. Due to the necessity of the components, we replace the GA and BO with a random search method with the same number of iterations, respectively. In addition, we also included an additional set of SCAR-style \cite{scar} mapping experiments. We migrate and adapt the mapping method in SCAR to the mapping representation of Compass to achieve a comparison with the SOTA multi-model mapping baseline. Fig. \ref{fig:ablation} shows the results of the ablation experiments, demonstrating the effectiveness of the specially designed GA and BO methods under the mapping representation of Compass.

\section{Conclusion}
In this work, we present Compass, a framework designed to address challenges posed by variable sequence lengths and mixed request types in LLM workloads on multi-chiplet accelerators. Compass introduces a computation execution graph-based mapping encoding scheme. Built on this foundation, Compass integrates an evaluation engine, a mapping generation engine based on GA, and a hardware sampling engine based on BO, enabling fast and flexible co-design of mapping and hardware. The solutions provided by Compass significantly reduce latency, energy consumption, and monetary cost. Furthermore, we analyze Compass in combination with cutting-edge inference service scheduling strategies, aiming to inspire future research and promote co-optimization of multi-chiplet accelerators and dynamic LLM workloads.

\section{Acknowledgement}


During the preparation of this manuscript, the authors used generative AI tools solely for language editing and grammar refinement. The AI tools were not used to generate scientific content, conduct analyses, interpret results, or formulate conclusions. All technical contributions, experimental results, and interpretations presented in this work were developed and verified by the authors. The authors take full responsibility for the accuracy, integrity, and originality of the content of this manuscript.

\bibliographystyle{IEEEtran}
\bibliography{ref}

\begin{IEEEbiography}[{\includegraphics[width=1in,height=1.25in,clip,keepaspectratio]{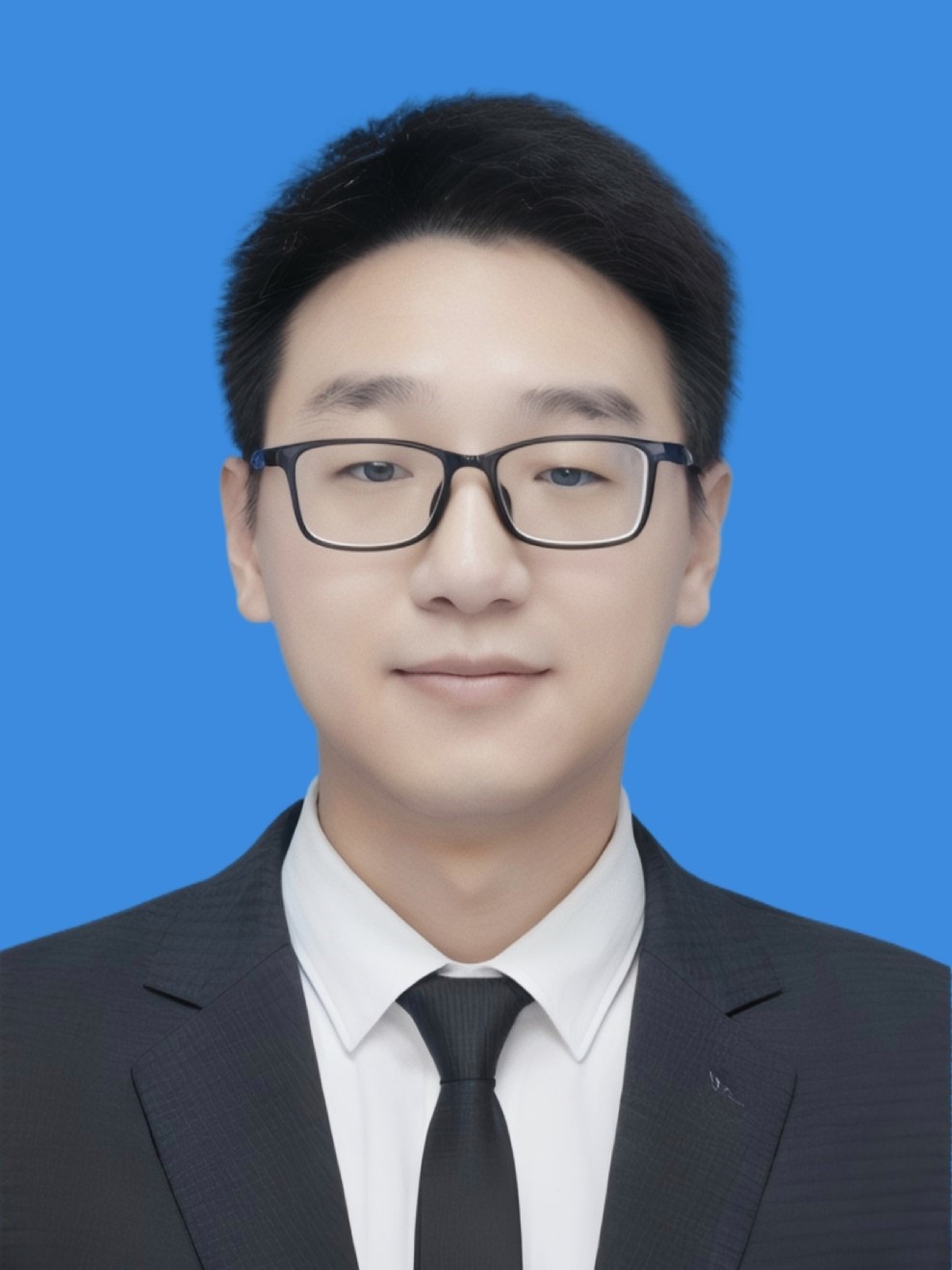}}]{Boyu Li}
received his B.S. degree in computer science and technology from Tongji University, in 2022. He is currently working toward the Eng.D degree in the School of Computer Science and Technology at the University of Science and Technology of China (USTC). His research interests include deep learning accelerator, multi-chiplet architecture, and inference optimization.\end{IEEEbiography}

\begin{IEEEbiography}[{\includegraphics[width=1in,height=1.25in,clip,keepaspectratio]{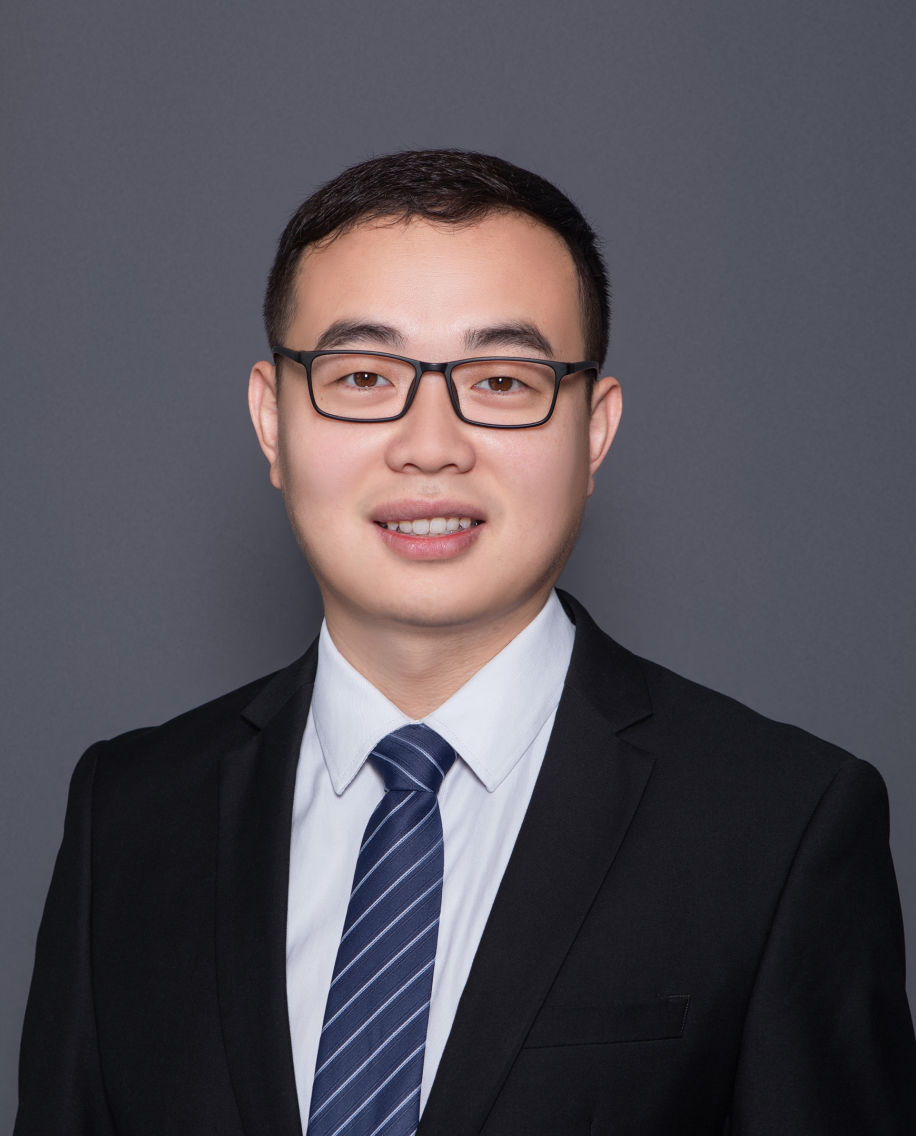}}]{Zongwei Zhu}
received his M.S. and Ph.D. degrees in Computer Science from the University of Science and Technology of China (USTC), in 2011 and 2014, respectively. From 2014 to 2016, he was a research assistant at the China University of Mining and Technology. From 2016 to 2018, he worked as a senior engineer at Huawei Company. Currently, he is an associate professor at the Suzhou Institute for Advanced Research of USTC. His research interests include AI architecture, edge computing, Low-power control and operating system.\end{IEEEbiography}

\begin{IEEEbiography}[{\includegraphics[width=1in,height=1.25in,clip,keepaspectratio]{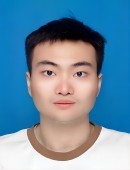}}]{Qianyue Cao} received the B.S. degree in computer science and technology from the Hefei University of Technology, Hefei, China, in 2021. He is currently pursuing Eng.D. in computer science at the School of Computer Science, University of Science and Technology of China, Hefei. His research focuses on edge computing and heterogeneous efficient computing.\end{IEEEbiography}

\begin{IEEEbiography}[{\includegraphics[width=1in,height=1.25in,clip,keepaspectratio]{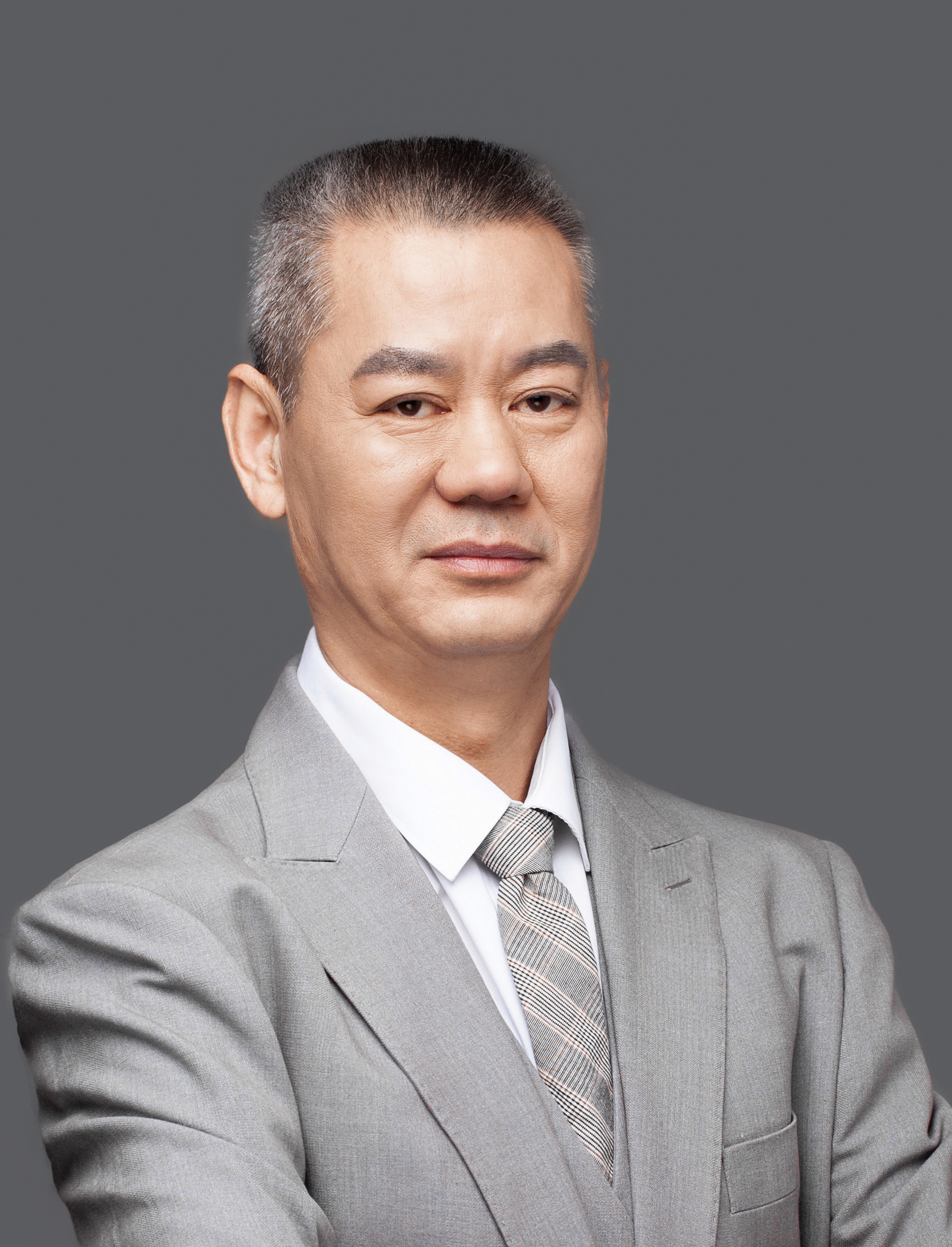}}]{Xi Li} received the Ph.D. degree in computer science from the University of Science and Technology of China, in 2003. He is currently a professor in the School of Computer Science and Technology, and the School of Software Engineering, University of Science and Technology of
China. There he directs the research programs in
High Energy-efficiency Intelligent Computing
Lab, examining various aspects of computer systems, especially real-time embedded systems.\end{IEEEbiography}

\begin{IEEEbiography}[{\includegraphics[width=1in,height=1.25in,clip,keepaspectratio]{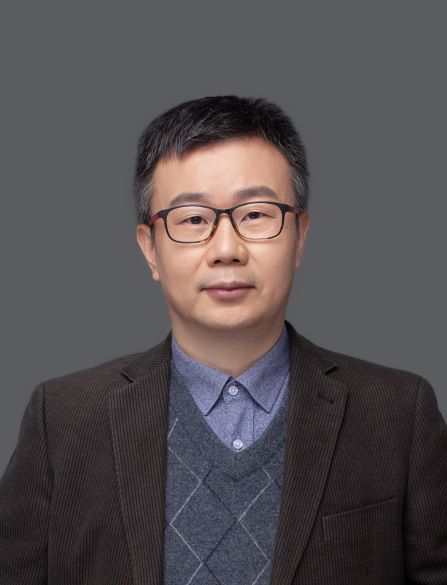}}]{Xuehai Zhou} (Member, IEEE)  received the B.S., M.S., and Ph.D. degrees in computer science from
the University of Science and Technology of China (USTC), Hefei, China, in 1987, 1990, and 1997, respectively. He is a Professor with the School of Computer Science and the School of Software Engineering, USTC. His research interest includes various aspects of multicore and distributed systems. Prof. Zhou serves as a General Secretary of the Steering Committee of Computer College Fundamental Lessons, and the Technical Committee of Open Systems, CCF.
\end{IEEEbiography}

\end{document}